\pdfoutput=1

\documentclass[11pt]{article}
\usepackage[table]{xcolor}
\usepackage{soul}
\usepackage{tcolorbox}
\usepackage{xcolor}

\definecolor{brightorange}{RGB}{255, 210, 100} 
\sethlcolor{brightorange}

\usepackage[final]{acl}

\usepackage{times}
\usepackage{latexsym}
\usepackage{booktabs}
\usepackage{amsmath}
\usepackage{subfigure}
\usepackage{colortbl} 
\usepackage{graphicx}

\usepackage[T1]{fontenc}

\usepackage[utf8]{inputenc}  

\usepackage{microtype}

\usepackage{inconsolata}

\usepackage{graphicx}

\usepackage{todonotes}

\title{Comprehensive Layer-wise Analysis of SSL Models \\ for Audio Deepfake Detection}

\author{Yassine El Kheir$^{1,2}$\thanks{Corresponding author: \texttt{yassine\_el\_kheir@dfki.de}}, Youness Samih$^{3}$, Suraj Maharjan, Tim Polzehl$^{1,2}$, Sebastian Möller$^{2,1}$ \\
  $^1$German Research Center for Artificial Intelligence (DFKI), Berlin, Germany \\
  $^2$Technical University of Berlin, Berlin, Germany \\
  $^3$IBM Research AI, Abu Dhabi, UAE \\
  \texttt{yassine\_el\_kheir@dfki.de, tim.polzehl@dfki.de, sebastian.moeller@dfki.de,} \\
  \texttt{younes.samih@ibm.com, mhjsuraj@gmail.com}
}

\begin{document}
\maketitle
\begin{abstract}

This paper conducts a comprehensive layer-wise analysis of self-supervised learning (SSL) models for audio deepfake detection across diverse contexts, including multilingual datasets (English, Chinese, Spanish), partial, song, and scene-based deepfake scenarios. By systematically evaluating the contributions of different transformer layers, we uncover critical insights into model behavior and performance. Our findings reveal that lower layers consistently provide the most discriminative features, while higher layers capture less relevant information. Notably, all models achieve competitive equal error rate (EER) scores even when employing a reduced number of layers. This indicates that we can reduce computational cost and increase the inference speed of detecting deepfakes by utilizing only a few lower layers. This work enhances our understanding of SSL models in deepfake detection, offering valuable insights applicable across varied linguistic and contextual settings. Our trained models and code are publicly available.\footnote{\url{https://github.com/Yaselley/SSL_Layerwise_Deepfake}}  

\end{abstract}

\section{Introduction}

Recent advancements in speech synthesis and voice conversion have resulted in high-fidelity synthetic audio, which can convincingly mimic real human voices \cite{kumar2023deep, huang2023singing, yi2023audio}. As a result, distinguishing between authentic speech and sophisticated deepfake audio has become increasingly challenging \cite{khan2022voice}. This poses a significant threat, particularly for Automatic Speaker Verification (ASV) systems which are widely being used for authenticating in access control, telephone banking, and forensic investigations \cite{anjum2017spoofing, li2024audio}. The potential for social and economic harm is evident, highlighted by many incidents like the \$243,000 scam, which exploited voice mimicry to deceive a CEO \cite{damiani2019voice}. Such threats emphasize the urgent need for effective audio deepfake detectors to ensure the security and trustworthiness of ASV systems.

In light of these challenges, SSL has emerged as a promising approach in speech processing. They are capable of extracting rich features from vast amounts of unlabeled audio data, which have enhanced the accuracy in speaker verification and speech recognition tasks~\cite{mohamed2022self, borgholt2022brief, yang2021superb, tsai2022superb, shon2022slue}. 
Many studies show that features derived from SSL methods consistently outperform traditional acoustic features across various tasks \cite{babu2021xls, xie2021siamese, martin2022vicomtech, wang2021investigating}. This suggests a new direction for enhancing the resilience of speech systems against deepfake threats.

While previous studies have highlighted the potential of SSL models in deepfake detection, a thorough layer-wise analysis of SSL models remains largely unexplored. Most current research concentrates on deepfakes created exclusively in English within full-utterance audio scenarios, leaving 
several important questions unanswered: (1) How effectively do SSL models detect deepfakes across diverse scenarios, including multiple languages, partial, song, and scene deepfakes? (2) How do different layers of SSL models behave in these varying conditions, and is their performance consistent across different setups? (3) Which layers yield the most discriminative features for differentiating real audio from deepfakes?

In this paper, we address these gaps by undertaking a comprehensive analysis of SSL models across various settings. This includes (1) full speech utterance deepfake detection in English (En), Chinese (Zh), and Spanish (Es); (2) partial speech utterance detection in English and Chinese; and (3) detection of songs and scene-based (acoustic environment) deepfakes. By examining layer-wise contributions, we aim to provide insights into the effectiveness of SSL models in multiple languages and contexts.

Our main contributions in this paper are as follows:
\begin{itemize}
    \item We show that the lower layers of the SSL models (Wav2Vec2, Hubert, and WavLM) are more important than the upper layers in detecting audio deepfakes. This is consistent across multiple languages and setups (full, partial deepfakes, song, scene) using different back-end classifiers.
    \item We demonstrate that utilizing only the lower layers of SSL models—4-6 layers for Small models and 10-12 layers for Large models—achieves performance comparable to, or sometimes better than, the full model while maintaining generalizability across diverse datasets and setups.
\end{itemize}
This research represents a pioneering effort to conduct a comprehensive layer-wise analysis of SSL models in the context of deepfake detection, encompassing a wide range of languages and settings.

\section{Background}

Traditional audio deepfake detection methods use hand-crafted acoustic features like MFCC, LFCC, or CQCC \cite{chakroborty2008improved, alegre2013one, li2023investigation, sahidullah15_interspeech, todisco2017constant}. These methods require expert knowledge and may overlook essential discriminative information \cite{tak2020explainability}. As a result, although it is effective on some datasets, such methods face challenges in generalizing to new or unseen spoofing attacks due to the static nature of the features. This limitation leaves systems vulnerable to emerging spoofing techniques. Therefore, there is a need for more robust and flexible anti-spoofing strategies that can automatically learn and extract relevant features from raw speech data.

With advancements in deep neural network (DNN) approaches, DNN-based methods have generally been adopted in two ways. First, they can be used as back-end models with traditional hand-crafted features as front-ends \cite{zhang122021effect}. In light of this,  light convolutional neural network (LCNN) \cite{lavrentyeva2017audio, tomilov2021stc}, residual neural network (ResNet) \cite{lai2019assert}, and siamese convolutional neural network \cite{lei2020siamese}, have been successfully applied to anti-spoofing systems using MFCC, LFCC, or CQCC features. On the other hand, end-to-end DNN modeling techniques has been used to extract learnable embeddings directly from raw audio and automatically extract features relevant to anti-spoofing techniques. For instance, in \cite{tak2021end}, SincNet \cite{ravanelli2018speaker} is utilized to directly extract front-end features from raw audio, with fixed cut-off frequencies. Other structures as well, as explored in \cite{bartusiak2022transformer, shim2022graph, khan2024frame}, are applied directly to raw audio data or combined with hand-crafted features to construct deep embeddings through supervised training.

SSL models are large pre-trained frameworks that serve as the backbone for various speech tasks. These models excel in capturing high-quality representations that can be fine-tuned for specific downstream applications. In the field of speech deepfake detection, this approach has been widely adopted, leading to state-of-the-art results. Studies have shown that fine-tuning SSL models like Wav2Vec2 \cite{baevski2020wav2vec} with a classifier can significantly improve detection performance \cite{tak2022automatic}. Alternatively, fine-tuning the XLS-R \cite{babu2021xls} model and using features from the fifth layer’s hidden states instead of the last layer has also proven effective \cite{lee2022representation}. Research highlights that utilizing hidden state features from various layers of pre-trained models can be highly beneficial for deepfake detection \cite{martin2022vicomtech}. 

Consequently, recent studies have focused on leveraging multi-layer features from SSL models to further enhance detection performance. Methods such as multi-fusion attentive classifiers \cite{guo2024audio}, attentive merging techniques \cite{pan2024attentive} with 10 \& 12 layers, and their improved versions \cite{guragain2024speech} have been explored. Additionally, expert fusion techniques, such as the Mixture of Experts method, have been proposed to extract and integrate relevant features for fake audio detection from multiple layers of SSL models \cite{wang2024mixture}.

\begin{figure*} 
\centering
\includegraphics[width=0.7\textwidth]{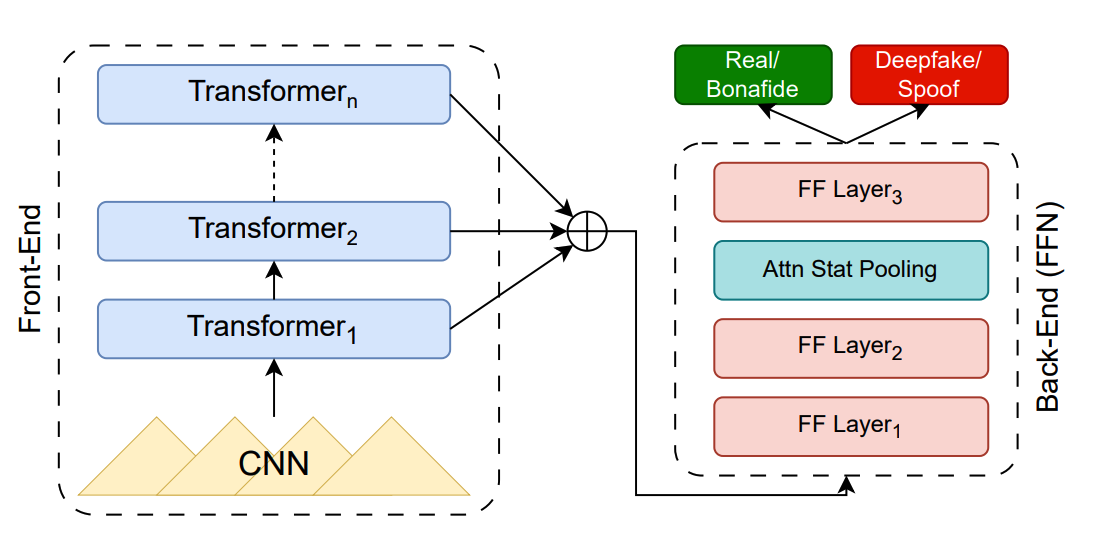}
\caption{\small Layer-wise Contribution Framework. The framework consists of SSL models as front-ends to extract features and a back-end classifier. The front-end SSL models remain frozen during the experiments to evaluate the layer-wise feature contribution.}
\label{fig:main_arch}
\end{figure*}

Recent advancements in multi-layer fusion techniques underscore the importance of a deeper examination of layer-wise contributions in SSL models. While previous research has largely focused on specific SSL models, particular classifiers, and predominantly English datasets \cite{lee2022representation, pan2024attentive}, our study offers a broader perspective.
Our methodology uniquely extends beyond these limitations by comprehensively assessing deepfake detection tasks across various SSL models, languages, and scenarios. This approach facilitates a more thorough analysis of layer-wise contributions.
Our detailed analysis reveals the most critical layers for audio deepfake detection, enabling the development of computationally efficient models that maintain effectiveness across diverse datasets. 

\section{Methodology}

Figure~\ref{fig:main_arch} shows our Layer-wise Contribution Framework. This framework helps us to understand the layer-wise contribution of different transformer layers in detecting deepfakes. We use SSL models as front-ends to extract features and a back-end classifier to perform classification (real vs deepfake). We freeze the SSL model and systematically investigate the contribution of each transformer layer to the final detection performance.

Let $\mathbf{x}$ represent the input audio signal, which is passed through an SSL model. This model is composed of $L$ transformer layers, where the output of each layer $l$ can be denoted as $\mathbf{h}_l$. The features $\mathbf{h}_l$ extracted from each layer $l \in \{1, 2, ..., L\}$ are then weighted by a layer-wise importance factor $w_l$.

We follow a weighted aggregation of the extracted features, where the final representation $\mathbf{h}_{\text{final}}$ used for classification is given by:

\begin{equation}
\mathbf{h}_{\text{final}} = \sum_{l=1}^{L} w_l \mathbf{h}_l
\label{eq:1}
\end{equation}

The weights $w_l$ are learnable parameters, and their values are constrained to sum to one by applying a softmax function:

\begin{equation}
  w_l = \frac{\exp(w_l)}{\sum_{k=1}^{L} \exp(w_k)} 
\label{eq:weights}
\end{equation}

This ensures that the contributions from different layers are normalized.
The back-end classifier receives $\mathbf{h}_{\text{final}}$ as input and generates the prediction $\hat{y}$ for the detection task. This setup allows us to investigate which layers contribute most to the detection of audio deepfakes across various settings.

\subsection{Front-Ends: SSL Models}

We experiment with 6 different front-end SSL models: small (base) and large versions of  Wav2Vec2, Hubert \cite{hsu2021hubert}, and  WavLM \cite{chen2022wavlm}. These models are the most commonly used SSL models for audio deepfake detection.  Wav2Vec2, Hubert, and WavLM also differ in their pre-training objectives.

\begin{figure*}
\includegraphics[width=1\textwidth]{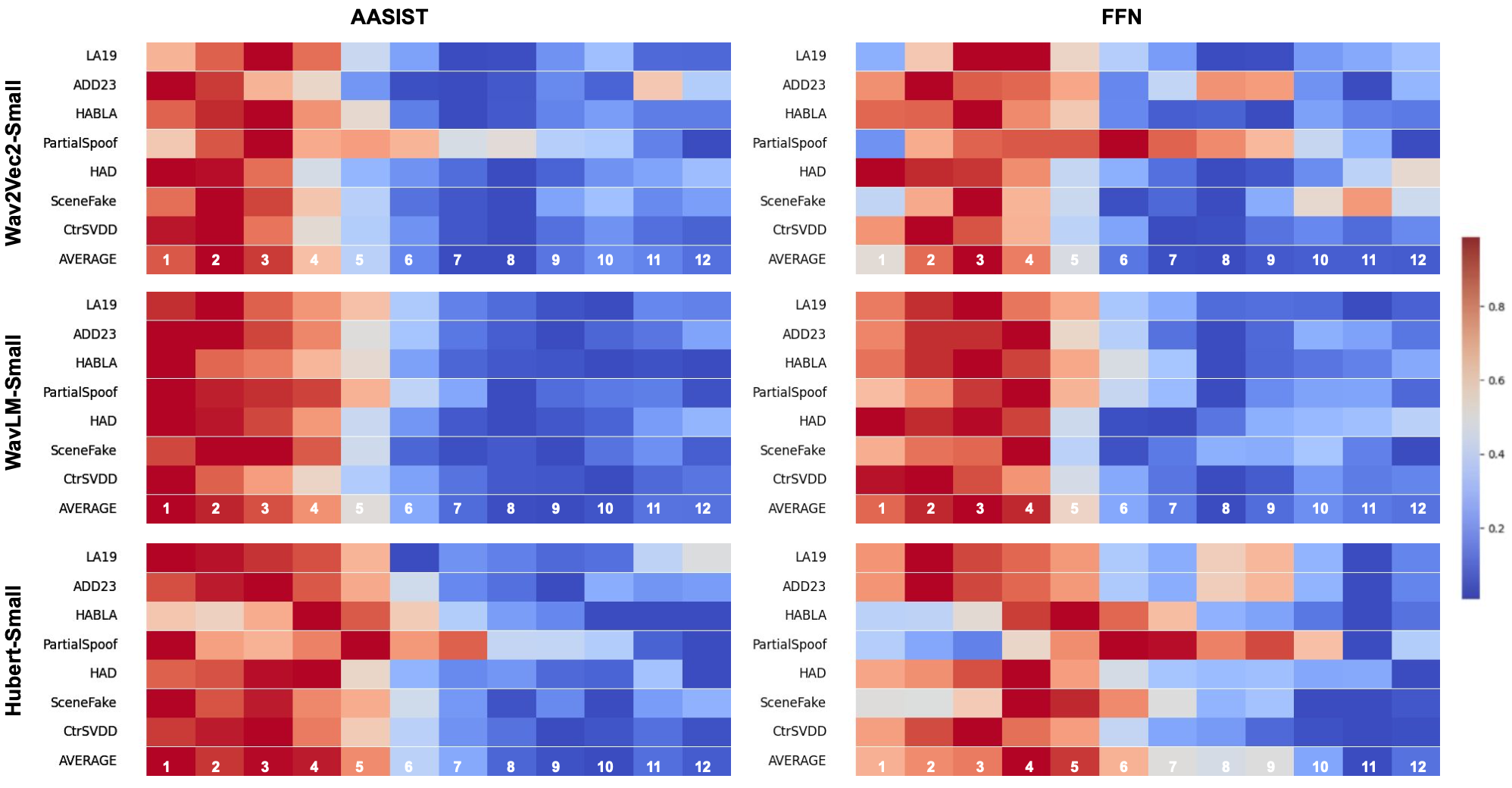}
\caption{Heatmap of Normalized Layer-wise weights Across Various Datasets using Small SSL Models. AVERAGE row representing the average weights across datasets.}
\label{fig:heat1}
\end{figure*}

\paragraph{Wav2Vec2.0}

Wav2Vec2 is a contrastive SSL model designed to learn speech representations by predicting latent features from masked audio regions. The architecture consists of two main components: a feature extractor and a stack of transformer layers. The feature extractor is a convolutional network (CNN) with 512 channels, designed to process audio segments of approximately 25ms from 16kHz sampled audio. The CNN compresses the input with strides and kernel sizes optimized to operate every 20ms. The model comes commonly in 2 configurations: the base version with 12 transformer layers and a hidden state dimension of 768, and the large version with 24 layers and a dimension of 1024. Wav2Vec2 is trained using Contrastive Predictive Coding (CPC) loss, which distinguishes between positive and negative samples by learning contextualized audio representations. This contrastive approach enables it to effectively model the underlying structure of speech data.

\paragraph{Hubert}

Hubert follows a similar architecture to Wav2Vec2.0, comprising a convolutional feature extractor followed by a stack of transformer layers. However, it differs in its self-supervised training objective. Instead of contrastive loss, Hubert employs a masked prediction approach, where the model is trained to predict discrete target representations derived from clustered speech features. Initially, these targets are obtained from k-means clustering on MFCC features, and in later stages, the model learns from its embeddings. This iterative refinement enables Hubert to capture meaningful speech representations without requiring frame-level labels. 

\paragraph{WavLM}

WavLM follows the same overall architecture as Wav2Vec2.0 and Hubert, consisting of a convolutional feature extractor and a stack of transformer layers. Like Hubert, it employs a masked prediction objective, but it extends this approach by incorporating speech enhancement techniques and training on both clean and noisy speech. This additional training data improves its robustness in real-world scenarios, particularly for speaker-related tasks such as speaker verification and separation. Compared to Hubert, WavLM introduces an auxiliary denoising task, enabling better performance in diverse acoustic environments. 

\paragraph{Model Selection and Multilingual Setup}

For our experiments, we select publicly available pre-trained models for both Wav2Vec2, Hubert, and WavLM. Specifically, we utilize multilingual versions of Hubert\footnote{\href{https://huggingface.co/utter-project/mHuBERT-147}{mHuBERT-147}}, and WavLM (both small and large)\footnote{\href{https://github.com/microsoft/unilm/blob/master/wavlm}{unilm/blob/master/wavlm}} and Wav2Vec2 Large\footnote{\href{https://github.com/facebookresearch/fairseq/blob/main/\\examples/wav2vec/xlsr}{fairseq/blob/main/examples/wav2vec/xlsr}}. For the Wav2Vec2 small model, we opt for the English pre-trained variant due to the absence of a small multilingual pre-trained version.

\subsection{Back-End Classifiers: FFN \textit{vs} AASIST}

We explore  two different back-end models: (1) a simple lightweight feedforward neural network (FFN), 
and (2) the state-of-the-art AASIST  \cite{jung2022aasist} model. We experiment with these two back-end models to demonstrate the generalizability of layer-wise contributions across different classifier architectures. This ensures that our findings are not confined to a specific back-end structure.

\begin{figure*}
\includegraphics[width=1\textwidth]{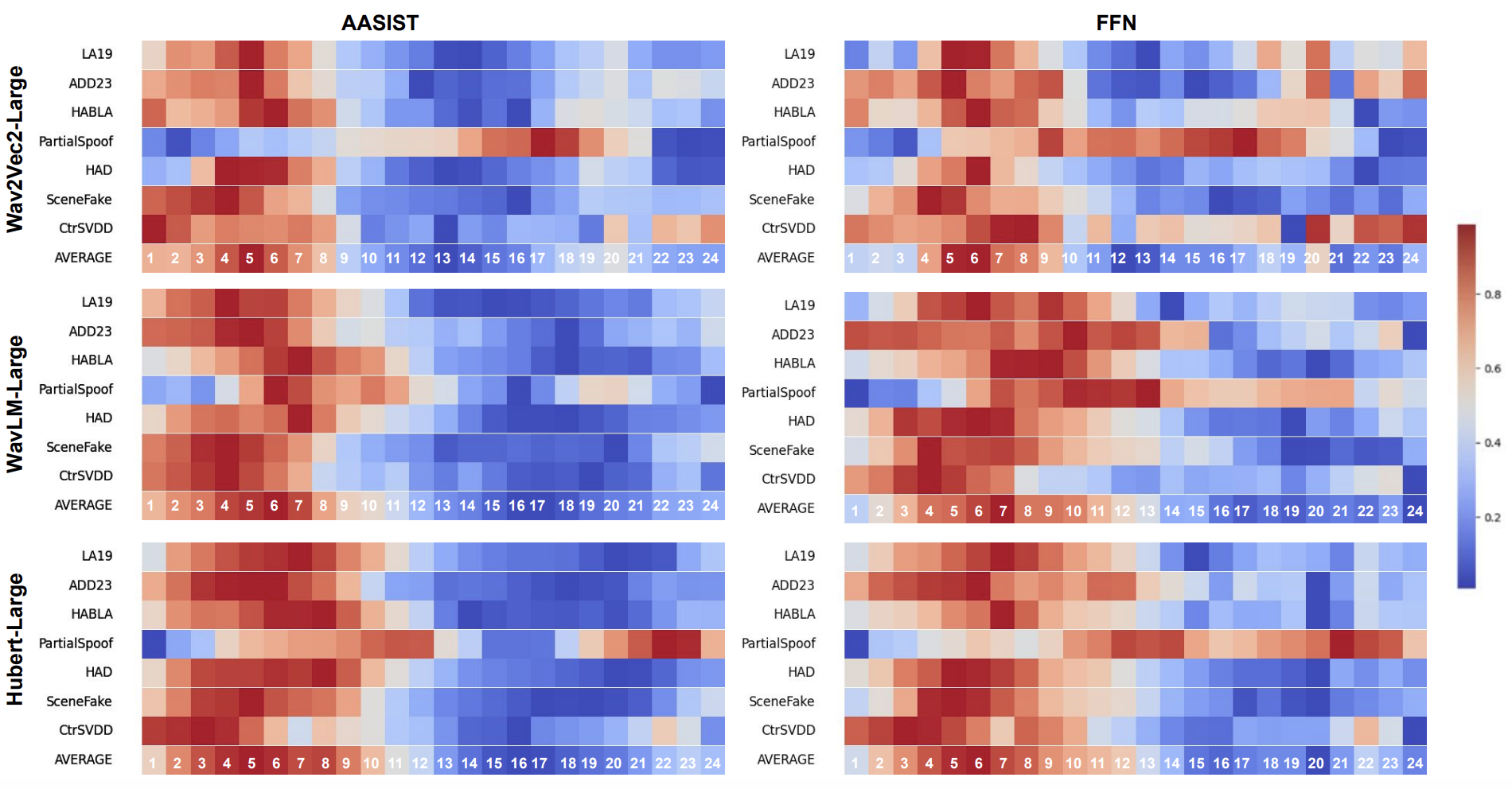}
\caption{Heatmap of Normalized Layer-wise weights Across Various Datasets using Large SSL Models. AVERAGE row representing the average weights across datasets.}
\label{fig:heat2}
\end{figure*}

\subsubsection{FFN}

We use the lightweight back-end classifier approach from \cite{martin2022vicomtech} for our deepfake detection task as shown in Figure \ref{fig:main_arch}.
Let $\mathbf{H} = [\mathbf{h}_1, \mathbf{h}_2, \dots, \mathbf{h}_T]$ represent the weighted sequence of feature vectors extracted from the front-end SSL model, as defined in Equation \ref{eq:1}. Each vector $\mathbf{h}_t \in \mathbf{R}^d$, where $d=768$ for SSL base models \& $d=1024$ for SSL large models, is first passed through a batch normalization layer. The normalized vectors are then processed by two feedforward (FF) layers with a hidden dimension of 128, using the SeLU activation function and dropout for regularization. Next, the processed vectors $\mathbf{h}_t$ are aggregated using an attentive statistical pooling layer \cite{okabe2018attentive}, which computes a weighted mean and standard deviation over the time steps, yielding a fixed-length representation for the entire utterance. This attentive pooling condenses the variable-length input sequence into a single vector of size $256$. Finally, the pooled representation is projected into $2$ classes.

\subsubsection{AASIST}

AASIST \cite{jung2022aasist} is a state-of-the-art model that utilizes graph-based attention mechanisms to capture both spectral and temporal audio features. Its key components include: (1) Graph Attention Layer (GAT), which computes attention maps for spectral and temporal features using linear layers, batch normalization, and SELU activation; (2) Heterogeneous Graph Attention Layer (HtrgGAT), which processes and refines both spectral and temporal nodes; (3) a graph pooling layer, which selects the top-k nodes based on attention scores; (4) residual blocks, which apply convolutional layers and SELU activation to process features; and (5) an attention mechanism that extracts spectral and temporal features from encoded data.

\section{Experimental Design}

\subsection{Datasets}

Table \ref{tab:datasets} summarizes datasets used in this study. These datasets are very diverse including both full utterance deep fakes (\textbf{Full Fake}) and partial utterance deepfakes (\textbf{Partial Fake}) in multiple languages: English (En), Chinese (Zh), and Spanish (Es). Additionally, \textbf{Scene}-based and \textbf{Song} deepfake setups are included. All datasets leverage cutting-edge TTS and VC generation techniques for creating audio deepfakes. Additional details on the different setups can be found in Appendix \ref{sec:setups}, while Appendix \ref{sec:datasets} provides further information on the datasets.

\begin{table*}[]
\centering
\scalebox{0.6}{
\begin{tabular}{lllcccccccccc}
\hline
                        &                    & \multicolumn{1}{l}{}                  & \multicolumn{10}{c}{\textbf{ Datasets}}                                                                                                                                                                                                                                                                                                                                                                          \\ \hline
\textbf{Back-End}       & \textbf{Front-end} & \multicolumn{1}{l}{\textbf{\#Layers}} & \textbf{LA19}                         & \textbf{LA21}                          & \textbf{DF21}                          & \textbf{ADD23.1.2.1}                   & \textbf{ADD23.1.2.2}                   & \textbf{HABLA}                         & \textbf{PartialSpoof}                 & \textbf{HAD\textsuperscript{\dag}}                           & \textbf{SceneFake}                    & \textbf{CtrSVDD}                       \\ \hline
\textbf{Wav2Vev2-Small} & \textbf{FFN}       & 2                                     & 0.63                                  & 6.31                                   & 15.09                                  & 58.31                                  & \cellcolor[HTML]{FFF2CC}\textbf{67.14} & \cellcolor[HTML]{FFF2CC}\textbf{9.80}  & 3.13                                  & 22.47                                  & 13.06                                 & 19.34                                  \\
                        &                    & 4                                     & 0.58                                  & 7.08                                   & \cellcolor[HTML]{FFF2CC}\textbf{12.1}  & \cellcolor[HTML]{FFF2CC}\textbf{60.07} & 69.73                                  & 9.87                                   & 2.64                                  & 24.18                                  & 11.20                                 & \cellcolor[HTML]{FFF2CC}\textbf{17.10} \\
                        &                    & 6                                     & \cellcolor[HTML]{FFF2CC}\textbf{0.48} & \cellcolor[HTML]{FFF2CC}\textbf{5.53}  & 14.02                                  & 63.30                                  & 73.77                                  & 9.87                                   & \cellcolor[HTML]{FFF2CC}\textbf{2.58} & 19.55                                  & 9.99                                  & 17.73                                  \\
                        &                    & 12                                    & 0.68                                  & 6.33                                   & 15.58                                  & 64.43                                  & 74.10                                  & 9.94                                   & 2.69                                  & \cellcolor[HTML]{FFF2CC}\textbf{17.63} & \cellcolor[HTML]{FFF2CC}\textbf{9.26} & 17.83                                  \\ \cline{2-13} 
                        & \textbf{AASIST}    & 2                                     & 0.63                                  & 5.09                                   & 15.01                                  & 62.59                                  & 66.85                                  & \cellcolor[HTML]{FFF2CC}\textbf{9.96}  & 2.97                                  & 37.43                                  & 12.48                                 & \cellcolor[HTML]{FFF2CC}\textbf{15.92} \\
                        &                    & 4                                     & \cellcolor[HTML]{FFF2CC}\textbf{0.55} & 5.69                                   & \cellcolor[HTML]{FFF2CC}\textbf{12.17} & \cellcolor[HTML]{FFF2CC}\textbf{56.12} & \cellcolor[HTML]{FFF2CC}\textbf{63.03} & 10.02                                  & \cellcolor[HTML]{FFF2CC}\textbf{1.82} & 36.41                                  & 9.87                                  & 20.42                                  \\
                        &                    & 6                                     & 0.92                                  & 6.48                                   & 12.50                                  & 57.78                                  & 65.62                                  & 10.09                                  & 2.11                                  & 34.65                                  & 10.87                                 & 16.80                                  \\
                        &                    & 12                                    & 1.02                                  & \cellcolor[HTML]{FFF2CC}\textbf{5.08}  & 15.80                                  & 58.59                                  & 64.91                                  & 10.11                                  & 2.25                                  & \cellcolor[HTML]{FFF2CC}\textbf{30.24} & \cellcolor[HTML]{FFF2CC}\textbf{9.54} & 16.67                                  \\ \hline
\textbf{WavLM-Small}    & \textbf{FFN}       & 2                                     & 3.41                                  & 27.13                                  & 17.98                                  & 57.27                                  & 63.22                                  & 9.91                                   & 3.36                                  & 31.32                                  & 12.27                                 & 19.18                                  \\
                        &                    & 4                                     & 2.55                                  & 30.15                                  & 13.90                                  & \cellcolor[HTML]{FFF2CC}\textbf{57.19} & 63.57                                  & \cellcolor[HTML]{FFF2CC}\textbf{9.88}  & 2.86                                  & 22.43                                  & 7.64                                  & 14.47                                  \\
                        &                    & 6                                     & \cellcolor[HTML]{FFF2CC}\textbf{1.85} & \cellcolor[HTML]{FFF2CC}\textbf{25.03} & \cellcolor[HTML]{FFF2CC}\textbf{12.34} & 57.46                                  & \cellcolor[HTML]{FFF2CC}\textbf{62.73} & 9.99                                   & \cellcolor[HTML]{FFF2CC}\textbf{2.12} & 21.74                                  & 4.07                                  & \cellcolor[HTML]{FFF2CC}\textbf{12.61} \\
                        &                    & 12                                    & 2.92                                  & 29.73                                  & 13.19                                  & 61.12                                  & 66.18                                  & 10.14                                  & 2.14                                  & \cellcolor[HTML]{FFF2CC}\textbf{19.00} & \cellcolor[HTML]{FFF2CC}\textbf{2.44} & 15.10                                  \\ \cline{2-13} 
                        & \textbf{AASIST}    & 2                                     & 0.55                                  & \cellcolor[HTML]{FFF2CC}\textbf{21.26} & 17.70                                  & 64.33                                  & 70.01                                  & 10.08                                  & 3.00                                  & 28.5                                   & 8.45                                  & 17.92                                  \\
                        &                    & 4                                     & \cellcolor[HTML]{FFF2CC}\textbf{0.26} & 21.51                                  & 12.42                                  & \cellcolor[HTML]{FFF2CC}\textbf{54.47} & \cellcolor[HTML]{FFF2CC}\textbf{58.86} & 10.14                                  & 2.18                                  & 31.03                                  & 5.79                                  & 15.89                                  \\
                        &                    & 6                                     & \cellcolor[HTML]{FFF2CC}\textbf{0.26} & 25.26                                  & \cellcolor[HTML]{FFF2CC}\textbf{11.76} & 55.85                                  & 60.02                                  & \cellcolor[HTML]{FFF2CC}\textbf{10.05} & \cellcolor[HTML]{FFF2CC}\textbf{1.84} & 30.87                                  & 3.80                                  & 14.83                                  \\
                        &                    & 12                                    & 0.57                                  & 39.87                                  & 15.48                                  & 58.55                                  & 66.05                                  & 10.10                                  & 1.97                                  & \cellcolor[HTML]{FFF2CC}\textbf{23.21} & \cellcolor[HTML]{FFF2CC}\textbf{2.68} & \cellcolor[HTML]{FFF2CC}\textbf{14.25} \\ \hline
\textbf{Hubert-Small}   & \textbf{FFN}       & 2                                     & 0.98                                  & 11.7                                   & 13.3                                   & \cellcolor[HTML]{FFF2CC}\textbf{56.46} & \cellcolor[HTML]{FFF2CC}\textbf{64.14} & 9.97                                   & 3.28                                  & 25.65                                  & 11.29                                 & 18.09                                  \\
                        &                    & 4                                     & 0.60                                  & 6.78                                   & 9.40                                   & 57.92                                  & 65.09                                  & 9.94                                   & 2.33                                  & 18.67                                  & 4.83                                  & 17.33                                  \\
                        &                    & 6                                     & \cellcolor[HTML]{FFF2CC}\textbf{0.44} & 6.92                                   & \cellcolor[HTML]{FFF2CC}\textbf{9.39}  & 59.81                                  & 67.63                                  & \cellcolor[HTML]{FFF2CC}\textbf{9.86}  & 2.54                                  & \cellcolor[HTML]{FFF2CC}\textbf{15.99} & \cellcolor[HTML]{FFF2CC}\textbf{3.25} & \cellcolor[HTML]{FFF2CC}\textbf{16.24} \\
                        &                    & 12                                    & 0.63                                  & \cellcolor[HTML]{FFF2CC}\textbf{6.61}  & 11.01                                  & 64.53                                  & 72.18                                  & 9.89                                   & \cellcolor[HTML]{FFF2CC}\textbf{2.09} & 18.49                                  & 4.33                                  & 16.51                                  \\ \cline{2-13} 
                        & \textbf{AASIST}    & 2                                     & 0.68                                  & 5.5                                    & 15.11                                  & 57.69                                  & \cellcolor[HTML]{FFF2CC}\textbf{63.17} & 10.03                                  & 3.37                                  & 32.62                                  & 8.70                                  & 17.73                                  \\
                        &                    & 4                                     & \cellcolor[HTML]{FFF2CC}\textbf{0.37} & 5.42                                   & 11.53                                  & \cellcolor[HTML]{FFF2CC}\textbf{58.10} & 66.38                                  & 10.04                                  & 2.05                                  & \cellcolor[HTML]{FFF2CC}\textbf{26.46} & \cellcolor[HTML]{FFF2CC}\textbf{4.31} & 14.72                                  \\
                        &                    & 6                                     & 0.39                                  & 4.55                                   & \cellcolor[HTML]{FFF2CC}\textbf{11.33} & 61.07                                  & 70.54                                  & \cellcolor[HTML]{FFF2CC}\textbf{10.01} & 1.96                                  & 29.72                                  & 4.52                                  & \cellcolor[HTML]{FFF2CC}\textbf{13.72} \\
                        &                    & 12                                    & 0.96                                  & \cellcolor[HTML]{FFF2CC}\textbf{4.35}  & 12.37                                  & 60.25                                  & 66.11                                  & 10.08                                  & \cellcolor[HTML]{FFF2CC}\textbf{1.88} & 29.65                                  & 6.31                                  & 18.73                                  \\ \hline
\end{tabular}}
\caption{Mean EER Results Across Datasets Using Full and Partial Transformer Layers of SSL Small Models. \# Indicates the Number of Layers Used from the Front-End SSL Models. Best Average EER is Highlighted in \hl{yellow}. \textsuperscript{\dag}We trained on the HAD training set and reported results on PartialSpoof eval set.}
\label{tab:tab1}
\end{table*}

\begin{table}[h!]
\centering
\scalebox{0.6}{
\begin{tabular}{lll}
\toprule
\textbf{Dataset} & \textbf{Language} & \textbf{Split} \\
\midrule
\multicolumn{3}{c}{\textbf{Full Fake}} \\
\midrule
ASVspoof 2019 (19LA) \cite{nautsch2021asvspoof} & En & Train \& Eval \\
ASVspoof 2021 (21LA)* \cite{yamagishi2021asvspoof} & En & Eval \\
ASVspoof 2021 (21DF)* \cite{yamagishi2021asvspoof} & En & Eval \\
ADD23 (Track 1.2)+ \cite{yi2023add} & Zh & Train \& Eval \\
HABLA \cite{florezhabla} & Es & Train \& Eval \\
\midrule
\multicolumn{3}{c}{\textbf{Partial Fake}} \\
\midrule
PartialSpoof \cite{zhang2021initial} & En & Train \& Eval \\
Half-Truth (HAD)\textsuperscript{\dag} \cite{yi2021half} & Zh & Train\\
\midrule
\multicolumn{3}{c}{\textbf{Song}} \\
\midrule
CtrSVDD \cite{zhang2024svdd} & Multilingual & Train \& Eval \\
\midrule
\multicolumn{3}{c}{\textbf{Scene}} \\
\midrule
SceneFake \cite{yi2024scenefake} & En & Train \& Eval \\
\bottomrule
\end{tabular}}
\caption{\small Summary of datasets used in this study. *Evaluated on models trained with LA19; + focused on Track 1.2 (Full Fake) with two eval sets, ADD23.1.2.1 and ADD23.1.2.2. \textsuperscript{\dag}We trained on the HAD training set and reported results on PartialSpoof eval set.}
\label{tab:datasets}
\end{table}

\begin{table*}[]
\centering
\scalebox{0.55}{
\begin{tabular}{lllcccccccccc}
\hline
                        &                    & \multicolumn{1}{l}{}                  & \multicolumn{10}{c}{\textbf{ Datasets}}                                                                                                                                                                                                                                                                                                                                                                        \\ \hline
\textbf{Back-End}       & \textbf{Front-end} & \multicolumn{1}{l}{\textbf{\#Layers}} & \textbf{LA19}                         & \textbf{LA21}                         & \textbf{DF21}                         & \textbf{ADD23.1.2.1}                   & \textbf{ADD23.1.2.2}                   & \textbf{HABLA}                         & \textbf{PartialSpoof}                 & \textbf{HAD\textsuperscript{\dag}}                           & \textbf{SceneFake}                    & \textbf{CtrSVDD}                       \\ \hline
\textbf{Wav2Vev2-Large} & \textbf{FFN}       & 4                                     & 0.37                                  & 19.77                                 & 14.45                                 & 55.59                                  & 69.45                                  & 9.92                                   & 4.43                                  & 35.19                                  & 9.09                                  & 17.65                                  \\
                        &                    & 6                                     & 0.18                                  & 13.35                                 & 7.74                                  & 51.08                                  & 65.51                                  & 9.89                                   & 3.81                                  & 24.89                                  & 4.02                                  & 14.19                                  \\
                        &                    & 8                                     & \cellcolor[HTML]{FFF2CC}\textbf{0.15} & 9.82                                  & 5.85                                  & 45.82                                  & 60.75                                  & \cellcolor[HTML]{FFF2CC}\textbf{9.91}  & 2.92                                  & 21.92                                  & 3.01                                  & 14.73                                  \\
                        &                    & 10                                    & 0.17                                  & 6.43                                  & 5.27                                  & 46.78                                  & 60.98                                  & 9.99                                   & 2.91                                  & 19.64                                  & \cellcolor[HTML]{FFF2CC}\textbf{2.73} & 13.61                                  \\
                        &                    & 12                                    & 0.21                                  & 5.96                                  & 4.20                                  & \cellcolor[HTML]{FFF2CC}\textbf{45.17} & \cellcolor[HTML]{FFF2CC}\textbf{57.76} & 9.96                                   & 2.91                                  & 18.42                                  & 3.65                                  & 14.31                                  \\
                        &                    & 24                                    & 0.21                                  & \cellcolor[HTML]{FFF2CC}\textbf{4.63} & \cellcolor[HTML]{FFF2CC}\textbf{4.14} & 46.29                                  & 59.91                                  & 10.01                                  & \cellcolor[HTML]{FFF2CC}\textbf{2.48} & \cellcolor[HTML]{FFF2CC}\textbf{16.83} & 2.96                                  & \cellcolor[HTML]{FFF2CC}\textbf{13.18} \\ \cline{2-13} 
                        & \textbf{AASIST}    & 4                                     & 0.69                                  & 13.41                                 & 18.77                                 & 61.76                                  & 70.28                                  & 10.05                                  & 8.11                                  & 43.73                                  & 6.40                                  & 19.3                                   \\
                        &                    & 6                                     & 0.54                                  & 9.61                                  & 12.40                                 & 52.83                                  & 66.10                                  & \textbf{10.04}                         & 5.78                                  & 33.27                                  & 4.15                                  & 15.04                                  \\
                        &                    & 8                                     & 0.90                                  & 11.05                                 & 9.80                                  & 46.56                                  & 61.11                                  & 10.05                                  & 3.56                                  & 30.58                                  & 4.38                                  & 14.37                                  \\
                        &                    & 10                                    & 0.80                                  & 8.65                                  & 10.51                                 & 46.57                                  & \cellcolor[HTML]{FFF2CC}\textbf{59.28} & 10.05                                  & 3.66                                  & 27.09                                  & \cellcolor[HTML]{FFF2CC}\textbf{3.08} & 14.12                                  \\
                        &                    & 12                                    & \cellcolor[HTML]{FFF2CC}\textbf{0.21} & \cellcolor[HTML]{FFF2CC}\textbf{7.95} & 9.38                                  & \cellcolor[HTML]{FFF2CC}\textbf{45.83} & 60.05                                  & \cellcolor[HTML]{FFF2CC}\textbf{10.04} & 2.99                                  & \cellcolor[HTML]{FFF2CC}\textbf{24.74} & 3.58                                  & 14.19                                  \\
                        &                    & 24                                    & 0.73                                  & 10.24                                 & \cellcolor[HTML]{FFF2CC}\textbf{8.12} & 49.69                                  & 61.31                                  & 10.11                                  & \cellcolor[HTML]{FFF2CC}\textbf{1.87} & 26.51                                  & 4.25                                  & \cellcolor[HTML]{FFF2CC}\textbf{13.34} \\ \hline
\textbf{WavLM-Large}    & \textbf{FFN}       & 4                                     & 1.32                                  & 13.17                                 & 15.97                                 & 57.85                                  & 69.87                                  & \textbf{9.99}                          & 5.99                                  & 27.05                                  & 5.14                                  & 21.27                                  \\
                        &                    & 6                                     & 0.43                                  & 8.82                                  & 12.36                                 & 55.01                                  & 68.51                                  & 10.03                                  & 5.05                                  & 24.93                                  & 4.89                                  & 19.88                                  \\
                        &                    & 8                                     & 0.33                                  & 8.00                                  & 9.32                                  & 52.11                                  & 68.02                                  & 10.03                                  & 3.57                                  & 19.02                                  & 4.19                                  & 16.62                                  \\
                        &                    & 10                                    & \cellcolor[HTML]{FFF2CC}\textbf{0.23} & 8.30                                  & 7.83                                  & 54.27                                  & 69.84                                  & 10.00                                  & 3.57                                  & 14.17                                  & 3.75                                  & 15.98                                  \\
                        &                    & 12                                    & 0.30                                  & \cellcolor[HTML]{FFF2CC}\textbf{4.95} & 8.29                                  & \cellcolor[HTML]{FFF2CC}\textbf{51.96} & \cellcolor[HTML]{FFF2CC}\textbf{67.98} & \cellcolor[HTML]{FFF2CC}\textbf{9.99}  & 3.32                                  & 13.59                                  & 3.27                                  & \cellcolor[HTML]{FFF2CC}\textbf{15.84} \\
                        &                    & 24                                    & 0.38                                  & 6.97                                  & \cellcolor[HTML]{FFF2CC}\textbf{6.24} & 56.54                                  & 68.19                                  & \cellcolor[HTML]{FFF2CC}\textbf{9.99}  & \cellcolor[HTML]{FFF2CC}\textbf{2.63} & \cellcolor[HTML]{FFF2CC}\textbf{9.90}  & \cellcolor[HTML]{FFF2CC}\textbf{3.17} & 16.84                                  \\ \cline{2-13} 
                        & \textbf{AASIST}    & 4                                     & 0.89                                  & 8.33                                  & 19.63                                 & 54.17                                  & 68.11                                  & \cellcolor[HTML]{FFF2CC}\textbf{10.06} & 12.85                                 & 39.79                                  & 3.61                                  & 17.25                                  \\
                        &                    & 6                                     & 0.44                                  & 10.14                                 & 13.88                                 & 55.05                                  & 67.69                                  & 10.11                                  & 8.45                                  & 41.48                                  & \cellcolor[HTML]{FFF2CC}\textbf{2.96} & 15.84                                  \\
                        &                    & 8                                     & \cellcolor[HTML]{FFF2CC}\textbf{0.38} & 8.04                                  & 12.72                                 & 53.57                                  & 68.03                                  & 10.11                                  & 7.96                                  & 37.36                                  & 3.31                                  & 16.14                                  \\
                        &                    & 10                                    & 0.57                                  & \cellcolor[HTML]{FFF2CC}\textbf{6.77} & 14.71                                 & 54.56                                  & 68.72                                  & 10.15                                  & 5.43                                  & \cellcolor[HTML]{FFF2CC}\textbf{36.21} & 3.49                                  & \cellcolor[HTML]{FFF2CC}\textbf{13.74} \\
                        &                    & 12                                    & 0.52                                  & 7.79                                  & 12.98                                 & \cellcolor[HTML]{FFF2CC}\textbf{48.50} & \cellcolor[HTML]{FFF2CC}\textbf{65.10} & 10.14                                  & 9.09                                  & 43.26                                  & 4.85                                  & 14.00                                  \\
                        &                    & 24                                    & 0.57                                  & 8.17                                  & \cellcolor[HTML]{FFF2CC}\textbf{9.29} & 51.21                                  & 67.32                                  & 10.12                                  & \cellcolor[HTML]{FFF2CC}\textbf{3.53} & 43.43                                  & 3.96                                  & 15.51                                  \\ \hline
\textbf{Hubert-Large}   & \textbf{FFN}       & 4                                     & 0.44                                  & 12.40                                 & 12.27                                 & 53.30                                  & 65.08                                  & \cellcolor[HTML]{FFF2CC}\textbf{9.96}  & 4.09                                  & 38.25                                  & 8.90                                  & 17.22                                  \\
                        &                    & 6                                     & \cellcolor[HTML]{FFF2CC}\textbf{0.19} & 10.34                                 & 8.69                                  & 52.77                                  & 65.82                                  & 9.89                                   & 3.63                                  & 29.58                                  & 4.71                                  & 14.67                                  \\
                        &                    & 8                                     & \cellcolor[HTML]{FFF2CC}\textbf{0.19} & 9.57                                  & 6.09                                  & 49.37                                  & 63.84                                  & 9.90                                   & 3.13                                  & 21.57                                  & 3.80                                  & 12.19                                  \\
                        &                    & 10                                    & 0.22                                  & 7.25                                  & 5.24                                  & \cellcolor[HTML]{FFF2CC}\textbf{48.61} & 62.74                                  & 9.89                                   & 2.81                                  & 16.20                                  & 4.02                                  & 12.49                                  \\
                        &                    & 12                                    & 0.20                                  & 5.70                                  & \cellcolor[HTML]{FFF2CC}\textbf{4.47} & 49.36                                  & \cellcolor[HTML]{FFF2CC}\textbf{62.43} & \cellcolor[HTML]{FFF2CC}\textbf{9.96}  & 2.57                                  & 15.59                                  & 3.34                                  & 12.71                                  \\
                        &                    & 24                                    & 0.28                                  & \cellcolor[HTML]{FFF2CC}\textbf{4.81} & 5.12                                  & 52.11                                  & 63.85                                  & \cellcolor[HTML]{FFF2CC}\textbf{9.95}  & \cellcolor[HTML]{FFF2CC}\textbf{2.33} & \cellcolor[HTML]{FFF2CC}\textbf{13.20} & \cellcolor[HTML]{FFF2CC}\textbf{2.97} & \cellcolor[HTML]{FFF2CC}\textbf{12.01} \\ \cline{2-13} 
                        & \textbf{AASIST}    & 4                                     & 0.70                                  & 12.26                                 & 16.98                                 & 53.29                                  & 63.53                                  & 10.06                                  & 6.54                                  & 34.97                                  & 4.79                                  & 17.18                                  \\
                        &                    & 6                                     & 0.64                                  & 10.58                                 & 15.07                                 & 48.54                                  & 61.47                                  & \cellcolor[HTML]{FFF2CC}\textbf{10.04} & 6.38                                  & 29.40                                  & 5.38                                  & 13.58                                  \\
                        &                    & 8                                     & 0.32                                  & 6.70                                  & 12.59                                 & \cellcolor[HTML]{FFF2CC}\textbf{45.35} & 58.91                                  & 10.08                                  & 4.19                                  & 29.45                                  & 4.14                                  & 11.17                                  \\
                        &                    & 10                                    & \cellcolor[HTML]{FFF2CC}\textbf{0.23} & 7.84                                  & 10.64                                 & 45.4                                   & 59.06                                  & 10.05                                  & 2.50                                  & 27.29                                  & 4.47                                  & \cellcolor[HTML]{FFF2CC}\textbf{10.52} \\
                        &                    & 12                                    & 0.27                                  & \cellcolor[HTML]{FFF2CC}\textbf{5.41} & 10.29                                 & 46.27                                  & \cellcolor[HTML]{FFF2CC}\textbf{56.97} & 10.06                                  & 2.47                                  & \cellcolor[HTML]{FFF2CC}\textbf{17.05} & \cellcolor[HTML]{FFF2CC}\textbf{3.11} & 11.04                                  \\
                        &                    & 24                                    & 0.47                                  & 6.65                                  & \cellcolor[HTML]{FFF2CC}\textbf{8.46} & 45.44                                  & 59.84                                  & 10.09                                  & \cellcolor[HTML]{FFF2CC}\textbf{2.37} & 22.67                                  & 3.64                                  & 11.16                                  \\ \hline
\end{tabular}}
\caption{Mean EER Results Across Datasets Using Full and Partial Transformer Layers of SSL Large Models. \# Indicates the Number of Layers Used from the Front-End SSL Models. Best Average EER is Highlighted in \hl{yellow}. \textsuperscript{\dag}We trained on the HAD training set and reported results on PartialSpoof eval set.}
\label{tab:tab2}
\end{table*}

\subsection{Training setup}
\label{sec:training}
We employ the cross-entropy (CE) loss function and the Adam optimizer, setting the learning rate to $1e^{-4}$ and a dropout rate of $0.2$, with a batch size of  $32$ for training our models.
During training, we freeze the SSL model parameters and update only the back-end classifier's parameters with the weights in Equation \ref{eq:1} initialized to ones.
We train our models for 50 epochs, applying early stopping with patience of 10 epochs based on evaluation loss. We use a single NVIDIA H100 GPU to run our experiments. Each experiment is repeated $3$ times with different seeds for reliable results and the average results are reported. We either crop or concatenate the audio data to create segments of approximately 4 seconds in duration (64,600 samples)\footnote{For \textbf{Partial Fake}, we used full audio}. We evaluate the model performance using the standard equal error rate (EER) metric. 

\section{Results and Discussion}

\subsection{Layer-wise Contribution Analysis}

After training SSL Model + Back-end classifiers with different datasets, we extract the layer-wise weights as outlined in Equation \ref{eq:weights} to study the contribution of each layer to the task.
For analysis, we first normalize these weights and then visualize them as heatmaps (we reported the average across 3 experiments for each dataset), as illustrated in Figures \ref{fig:heat1} and \ref{fig:heat2}.

\paragraph{Small SSL Models}

Figure \ref{fig:heat1} presents the normalized layer-wise weights after training Wav2Vec2-Small, Hubert-Small, and WavLM-Small front-end models, using FFN and AASIST back-ends across various datasets. The results consistently show that the highest contributions come from lower layers (1-6) across all datasets and setups (speech, scene, and song), indicating that these layers capture the most discriminative features for the deepfake detection task. This pattern holds regardless of the back-end classifier used, suggesting that the classifier choice does not significantly impact the task reliance on lower layers. However, slight variations are observed in certain cases, such as PartialSpoof where intermediate layers contribute more significantly using FFN.

Looking at the average (AVERAGE) layer-wise weights across datasets, the trend becomes even clearer, with lower layers (1-4) consistently showing the highest contributions. This confirms the dominant role of lower layers in the deepfake performance of small SSL models.

\paragraph{Large SSL Models} Figure \ref{fig:heat2} shows the normalized layer-wise weights for Wav2Vec2-Large, Hubert-Large, and WavLM-Large front-end models. Unlike the concentrated lower-layer contributions seen with small SSL models, large SSL models display a lesser concentrated pattern of layer importance, especially with the  FFN back-end. For Wav2Vec2-Large, Hubert-Large, and WavLM-Large, lower layers (2-13) still contribute significantly across most datasets when using both FFN and AASIST back-ends. However, a shift towards middle layers (12-21) is observed in PartialSpoof. When looking at the average (AVERAGE) layer-wise weights, this trend remains stable across SSL models and back-end classifiers, with the highest contributions emerging in lower layers (1-12), particularly around the 4-7th layers.

\begin{table*}[]
\centering
\scalebox{0.7}{
\begin{tabular}{lccc|lccc|lccc|c}
\hline
\textbf{}         & \multicolumn{3}{c|}{\textbf{Wav2Vec-Small}}       & \textbf{}         & \multicolumn{3}{c|}{\textbf{WavLM-Small}}         & \textbf{}         & \multicolumn{3}{c|}{\textbf{Hubert-Small}}        & \textbf{AVG\(_{out}\)}                  \\ \hline
\textbf{\#Layers} & \textbf{FFN}   & \textbf{AASIST} & \textbf{AVG\(_{in}\)}   & \textbf{\#Layers} & \textbf{FFN}   & \textbf{AASIST} & \textbf{AVG\(_{in}\)}   & \textbf{\#Layers} & \textbf{FFN}   & \textbf{AASIST} & \textbf{AVG\(_{in}\)}   & \multicolumn{1}{l}{\textbf{}} \\ \hline
\textbf{2}        & 21.53          & 22.89           & 22.21          & \textbf{2}        & 24.51          & 24.18           & 24.35          & \textbf{2}                 & 21.49          & 21.46           & 21.48          & 22.68                         \\
\textbf{4}        & \textbf{21.46} & 21.61           & \textbf{21.54} & \textbf{4}        & 22.46          & \textbf{21.26}  & 21.86          & \textbf{4}                 & 19.29          & \textbf{19.94}  & \textbf{19.62} & 21.01                         \\
\textbf{6}        & 21.68          & 21.78           & 21.73          & \textbf{6}        & \textbf{20.99} & 21.45           & \textbf{21.22} & \textbf{6}                 & \textbf{19.21} & 20.78           & 20.00             & \textbf{20.98}                \\
\textbf{12}       & 21.85          & \textbf{21.42}  & 21.64          & \textbf{12}       & 22.20           & 23.27           & 22.74          & \textbf{12}                & 20.63          & 21.07           & 20.85          & 21.74                         \\ \hline
                  & \multicolumn{3}{c|}{\textbf{Wav2Vec-Large}}       &                   & \multicolumn{3}{c|}{\textbf{WavLM-Large}}         & \textbf{}         & \multicolumn{3}{c|}{\textbf{Hubert-Large}}        & \multicolumn{1}{l}{\textbf{}} \\ \hline
\textbf{\#Layers} & \textbf{FFN}   & \textbf{AASIST} & \textbf{AVG\(_{in}\)}   & \#Layers          & \textbf{FFN}   & \textbf{AASIST} & \textbf{AVG\(_{in}\)}   & \textbf{\#Layers} & \textbf{FFN}   & \textbf{AASIST} & \textbf{AVG\(_{in}\)}   & \multicolumn{1}{l}{\textbf{}} \\ \hline
\textbf{4}        & 23.59          & 25.25           & 24.42          & \textbf{4}        & 22.76          & 23.47           & 23.12          & \textbf{4}                 & 22.19          & 22.03           & 22.11          & 23.22                         \\
\textbf{6}        & 19.47          & 20.98           & 20.23          & \textbf{6}        & 20.99          & 22.6            & 21.80           & \textbf{6}                 & 20.03          & 20.11           & 20.07          & 20.70                          \\
\textbf{8}        & 17.49          & 19.24           & 18.37          & \textbf{8}        & 19.12          & 21.76           & 20.44          & \textbf{8}                 & 17.97          & 18.29           & 18.13          & 18.98                         \\
\textbf{10}       & 16.85          & 18.38           & 17.62          & \textbf{10}       & 18.79          & 21.44           & 20.12          & \textbf{10}                & 16.95          & 17.80            & 17.38          & 18.37                         \\
\textbf{12}       & 16.26          & \textbf{17.90}   & \textbf{17.08} & \textbf{12}       & \textbf{17.95} & 21.62           & 19.79          & \textbf{12}                & \textbf{16.63} & \textbf{16.29}  & \textbf{16.46} & \textbf{17.78}                \\
\textbf{24}       & \textbf{16.06} & 18.62           & 17.34          & \textbf{24}       & 18.09          & \textbf{21.31}  & \textbf{19.7}  & \textbf{24}                & 16.66          & 17.08           & 16.87          & 17.97                         \\ \hline
\end{tabular}
}
\caption{Reported average EER across datasets using different Wav2Vec2, Hubert, and WavLM SSL models with different back-end layers. \text{AVG\(_{in}\)} represents the average EER across datasets for each SSL model, calculated by first averaging the EER for different back-ends (FFN and AASIST) within the same reduced-layer model. \text{AVG\(_{out}\)} indicates the overall average of \text{AVG\(_{in}\)}, calculated separately for SSL Small and SSL Large models. \#L indicates the number of transformer layers.
}
\label{tab:final_one}
\end{table*}


\paragraph{Focus on Local Features Across Models}
Across all models, there is a consistent reliance on the lower layers, particularly layers 1-6 in Small SSL models and layers 1-12 in Large SSL models. This trend suggests that the models focus on local features, which aligns with how audio deepfake detectors identify artifacts left by VC and TTS algorithms. These artifacts often reside in specific frequency sub-bands or temporal segments, highlighting the significance of local feature extraction \cite{sriskandaraja2016investigation, tak2020explainability}. 


\noindent Particularly, in our investigation of the PartialSpoof data, we observed significant disfluency in certain speech segments, indicating that the model tends to focus more on the transitions between fake and real segments as shown in \cite{liu2024neural}. This may explain the model's increased reliance on intermediate and upper SSL layers, which are more effective at capturing disfluent speech patterns, as demonstrated in \cite{shih2024self}.

\subsection{Performance of Reduced-Layer SSL Models}

This section investigates the capabilities of SSL models when limited to a reduced number of transformer layers. We evaluate Wav2Vec2-Small, Hubert-Small, and WavLM-Small using 2, 4, and 6 layers, as well as Wav2Vec2-Large, Hubert-Large, and WavLM-Large using 4, 6, 8, 10, and 12 layers. These configurations are guided by our earlier layer-wise analysis. The reduced-layer models are trained under the same conditions as the initial full models described in Section \ref{sec:training}, and their performance is assessed across the same datasets and setups. We introduce a notation where "SSLModel-X" represents a model with X layers (e.g., Wav2Vec2-Small-4 denotes Wav2Vec2-Small with 4 layers).

\paragraph{Performance Comparison Across Deepfake Scenarios}

Our analysis demonstrates that reduced-layer SSL models perform on par with, or even better than, full-layer models in various deepfake detection scenarios. As shown in Tables \ref{tab:tab1} and \ref{tab:tab2}, no single configuration consistently outperforms across all datasets and back-end classifiers. While full models excel in some cases, reduced models—especially 4- and 6-layer configurations for small models and 8-, 10-, and 12-layer configurations for large models—achieve comparable and superior results in others.

For example, Wav2Vec2-Small-6+FFN achieves a lower EER of \textbf{0.48\%} on LA19 compared to \textbf{0.68\%} for the full model, with better generalization on LA21 and DF21. WavLM-Small-6+FFN outperforms the full model on 6 out of 10 datasets. The WavLM-Small-4+AASIST configuration yields an EER of \textbf{54.47\%} on ADD23.1.2.1, compared to \textbf{58.55\%} for the full model, indicating the effectiveness of reduced layers. Hubert-Small shows similar trends, with reduced-layer models performing either comparably or better than the full model. For instance, the Hubert-Small-6+ASSIST model scores \textbf{13.72\%} on the CtrSVDD dataset, outperforming the full model's \textbf{18.73\%}. However, on the SceneFake datasets, full models for both Wav2Vec-Small and WavLM-Small consistently outperform the reduced-layer versions, although the difference is minimal, ranging from \textbf{0.5\%} to \textbf{1.5\%} EER.

For Large SSL models, the Wav2Vec2-Large-12+AASIST configuration significantly reduces the EER to \textbf{0.21\%} on LA19, outperforming the full model while maintaining as well generalization on LA21 and DF21. Specifically, it shows a \textbf{2.29\%} gain on LA21, with only a slight \textbf{1.26\%} reduction on DF21. The WavLM-Large-12+FFN configuration achieves \textbf{15.84\%} EER on CtrSVDD, outperforming the full model. Hubert-Large follows a similar trend, with the 8-layer model reaching \textbf{0.19\%} EER on LA19, and Hubert-Large-12+FFN achieving the best result on DF21 with an EER of \textbf{4.47\%}. 

Across critical datasets such as ADD23, reduced-layer models consistently outperform full models: Wav2Vec2-Large-12+FFN achieves \textbf{45.17\%} EER on ADD23.1.2.1, WavLM-Large-12+FFN reaches \textbf{51.96\%}, and Hubert-Large-10 hits \textbf{48.61\%}. In contrast, for PartialSpoof, full models consistently outperform the reduced-layer SSL models, aligning with our observation in Figure \ref{fig:heat2}, where the upper-middle layers contribute most to this dataset. Nonetheless, despite the reduced-layer SSL models not significantly lowering the results, the differences between the best reduced-layer models and full models remain minimal, with EER differences ranging from \textbf{0.24\%} to \textbf{1.9\%}.

\subsection{Optimal Layer Configuration Across Datasets}


Our results reveal that using 4-6 layers for the Small models (Wav2Vec-Small, Hubert-Small, WavLM-Small) and 10-12 layers for the Large models (Wav2Vec2-Large, Hubert-Large, WavLM-Large) consistently delivers strong average performance across various datasets. As illustrated in Table \ref{tab:final_one}, \textbf{AVG\(_{in}\)} represents the average EER across datasets for each SSL model, computed by averaging the EER for the different back-end configurations (FFN and AASIST) within the same reduced-layer model. The \textbf{AVG\(_{out}\)} column provides the overall average of \textbf{AVG\(_{in}\)}, calculated separately for Small and Large SSL models. Notably, Wav2Vec2-Small-6, Hubert-Small-6, and WavLM-Small-6 achieve an \textbf{AVG\(_{out}\)} of \textbf{20.98\%}, while Wav2Vec2-Small-4, Hubert-Small-4, and WavLM-Small-4 reach \textbf{21.01\%}, outperforming both the full SSL model and other reduced-layer configurations. Likewise, the 12-layer configuration for the Large models also demonstrates optimal performance, with an \textbf{AVG\(_{out}\)} of \textbf{17.78\%}, showing its robustness across datasets. However, in some cases, the full model outperforms other configurations, such as with Wav2Vec-Small-AASIST, Wav2Vec2-Large-FFN, and WavLM-Large-AASIST. Despite this, the reduced-layer models come close in performance, with minimal degradation in EER, ranging from a \textbf{0.31\%} to \textbf{0.72\%} difference.

\section{Small \textit{vs} Large SSL Models}
When comparing Small and Large SSL models across various datasets, no significant performance difference is observed for most datasets, including LA19, LA21, HABLA, PartialSpoof, and SceneFake. However, for more recent datasets, where deepfake models are more advanced, such as DF21, ADD23.1.2.1, and ADD23.1.2.2, larger architectures are essential for capturing subtle deepfake artifacts and improving generalization across languages, as demonstrated with HAD\footnote{Models trained on HAD training set but evaluated on PartialSpoof eval set.}. For instance, Tables~\ref{tab:tab1} and \ref{tab:tab2} show a \textbf{10.95\%} difference in EER is observed when comparing the best performance of Wav2Vec-Small on ADD23.1.2.1 with Wav2Vec-Large. A \textbf{6.2\%} difference in EER is also noted between WavLM-Small and WavLM-Large on ADD23.1.2.2. On the other hand, Hubert-Large achieves its best performance on CtrSVDD with an EER of \textbf{10.53\%}, lagging behind the best Hubert-Small configuration by \textbf{3.19\%} EER difference.

\section{Conclusion}
This study provides a comprehensive layer-wise analysis of SSL models for audio deepfake detection, covering multilingual datasets and a range of deepfake types, including Full Fake, Partial Fake, and Song- and Scene-based scenarios. 
Our results show that using only first 4-6 layers for Small models and 10-12 layers for Large models consistently delivers optimal performance across datasets, reducing parameters by at least half. This significantly lowers computational costs while maintaining competitive EER scores.

\section{Limitations}

Our study provides a comprehensive layer-wise analysis of SSL models for deepfake detection across languages and contexts, but it has some limitations. We did not investigate the latest Audio Language Models and Speech tokenizers used for audio deepfake generation \cite{xie2024codecfake}, which might lead to different artifacts and model behavior. Additionally, we focused solely on speech-based SSL models, excluding acoustic SSL models. Future work could explore these models, especially for tasks involving non-speech artifacts, like those in the SceneFake dataset. We also plan to investigate alternative fusion strategies, such as attentional fusion, beyond a simple weighted-sum layer-wise combination.

\bibliography{custom}

\appendix
\section{Setup Description}
\label{sec:setups}
In this study, various types of audio deepfakes are utilized to evaluate the performance of SSL. These deepfakes are categorized as follows:

\textbf{Full Fake:} The entire utterance is generated using voice conversion (VC) or text-to-speech (TTS) technologies. No part of the original speech remains intact.

\textbf{Partial Fake:} Only specific segments or words from the original utterance are replaced with VC or TTS-generated content. The remainder of the utterance is real.

\textbf{Song:} These setups involve utterances accompanied by background music, with corresponding speech content either real or deepfaked.

\textbf{Scene:} In these cases, the speech itself is real, but the background environment or context has been altered (e.g., a different ambiance or noise is introduced to simulate a new scene).

\section{Datasets}
\label{sec:datasets}
In this section, we provide information about different datasets used in the study: 

\subsection{ASVSpoof19 (LA19):}

Our work focuses on the LA subset of the ASVSpoof19 database, which contains both bonafide (real) and spoofed speech (deepfake) data. The spoofed data is generated using 17 different TTS and Voice VC systems. While the TTS and VC systems were trained on data from the VCTK\footnote{https://datashare.ed.ac.uk/handle/10283/2651} database, there is no overlap with the data in the 2019 database. Of the 17 systems, 6 are categorized as known attacks, and the remaining 11 are classified as unknown. The training and development sets consist exclusively of known attacks, while the evaluation set includes 2 known attacks and 11 unknown spoofing attacks. Among the 6 known attacks, 2 are VC systems and 4 are TTS systems.

\subsection{ASVSpoof21 (LA21 \& DF21):}

The ASVspoof21 evaluation dataset contains bonafide and spoofed speech transmitted over various telephony systems, including VoIP and PSTN. While no additional noise is introduced, transmission introduces variability and artifacts from both spoofing and encoding. Unlike previous LA tasks, LA21 includes acoustic propagation, with all data featuring reverberation and noise. The evaluation set includes recordings from the same 48 speakers (21 male, 27 female) as ASVspoof 2019. The DF21 task simulates an attack where the attacker has access to a victim’s voice data, such as audio from social media. The dataset is sourced from the VCTK corpus and additional undisclosed datasets.

\subsection{ADD23 (Track1.2):}
ADD23 Track 1.2 focuses on detecting fake utterances, particularly those generated in Track 1.1 (from the ADD23 challenge). The training and development sets are the same as those used in Track 3.2 of ADD 2022\cite{yi2022add}, comprising both real and fake utterances based on the AISHELL-3 dataset \cite{shi2020aishell}. The evaluation phase includes two distinct datasets, referred to as ADD23.1.2.1 and ADD23.1.2.2.

\subsection{HABLA:}

The HABLA dataset is a Spanish-language anti-spoofing corpus that includes accents from Argentina, Colombia, Peru, Venezuela, and Chile. It contains more than 22,000 genuine speech samples from both male and female speakers across these five countries, alongside 58,000 spoofed samples generated using six different speech synthesis methods. We use the train/dev/eval split provided by the dataset authors.

\subsection{PartialSpoof:}

The PartialSpoof database is derived from the ASVspoof 2019 LA dataset, which includes 17 types of spoofed data generated by advanced speech synthesizers, voice converters, and hybrid systems. The same bonafide data from the ASVspoof19 LA corpus is used to create partially spoofed audio by following these steps: first, voice activity detection (VAD) algorithms are applied, and based on the detected boundaries, a randomly selected segment of a bonafide utterance are replaced with a spoofed segment.

\subsection{Half-Truth (HAD):}

The HAD (Half-Truth Audio Detection) dataset is focused on detecting partially fake audio, where only a small portion of an utterance—such as individual words—is altered using advanced TTS systems. It is based on the AISHELL-3 speech corpus, which includes multi-speaker recordings designed for TTS model training. The dataset provides the challenging task of identifying subtle manipulations in audio where most of the content remains genuine.

\subsection{CtrSVDD:}

CtrSVDD is a diverse dataset focused on detecting deepfake singing vocals. It consists of 47.64 hours of bona fide singing and 260.34 hours of deepfake vocals generated using state-of-the-art synthesis techniques. Deepfake songs span 14 different methods, including 7 speech synthesis and 7 voice conversion (VC) techniques. The dataset includes 164 singer identities and provides a total of 188,486 deepfake song clips and 32,312 genuine song clips, with an average clip length of 5.02 seconds. CtrSVDD offers a rich resource for studying deepfake detection in singing, making it a valuable tool for advancing research in this area.

\subsection{SceneFake:}

SceneFake is a dataset developed for scene fake audio detection, focusing on manipulated audio samples generated by altering the acoustic scene of real utterances using advanced speech enhancement technologies. Unlike traditional datasets that modify timbre, prosody, or linguistic content, SceneFake specifically addresses the challenge of detecting audio where the background environment has been tampered with. The dataset includes a variety of real utterances paired with their manipulated counterparts, and initial experiments indicate that existing models trained on datasets like ASVSpoof19 struggle to reliably detect scene fake utterances.

\end{document}